\documentclass[prd,onecolumn,nofootinbib]{revtex4}
\usepackage{bm,amsmath,amssymb,graphicx}

\begin{document}

\noindent
The Astrophysical Journal {\bf 832} (2), 96 (2016)\\

\title{Universe in a black hole in Einstein--Cartan gravity}
\author{Nikodem Pop{\l}awski}
\altaffiliation{NPoplawski@newhaven.edu}
\affiliation{Department of Mathematics and Physics, University of New Haven, West Haven, CT, USA}

\begin{abstract}
The conservation law for the angular momentum in curved spacetime, consistent with relativistic quantum mechanics, requires that the antisymmetric part of the affine connection (torsion tensor) is a variable in the principle of least action.
The coupling between the spin of elementary particles and torsion in the Einstein--Cartan theory of gravity generates gravitational repulsion at extremely high densities in fermionic matter, approximated as a spin fluid, and thus avoids the formation of singularities in black holes.
The collapsing matter in a black hole should therefore bounce at a finite density and then expand into a new region of space on the other side of the event horizon, which may be regarded as a nonsingular, closed universe.
We show that quantum particle production caused by an extremely high curvature near a bounce can create enormous amounts of matter, produce entropy, and generate a finite period of exponential expansion (inflation) of this universe.
This scenario can thus explain inflation without a scalar field and reheating.
We show that, depending on the particle production rate, such a universe may undergo several nonsingular bounces until it has enough matter to reach a size at which the cosmological constant starts cosmic acceleration.
The last bounce can be regarded as the big bang of this universe.\\ \\
Key words: black hole physics, cosmology: theory, early universe, gravitation, inflation.
\end{abstract}
\maketitle

{\bf 1. Introduction}\\ \\
Relativistic quantum mechanics predicts that elementary particles that are fermions, described by the Dirac equation, have the intrinsic angular momentum (spin).
The conservation law for the total angular momentum (orbital plus intrinsic) that admits the exchange between its orbital and intrinsic components (spin--orbit interaction) in curved spacetime requires an asymmetric affine connection \cite{req}.
The Einstein--Cartan or Einstein--Cartan--Sciama--Kibble (ECSK) theory of gravity naturally extends the metric general relativity by removing its constraint of the symmetry of the connection \cite{ScKi}.
The antisymmetric part of the affine connection, the torsion tensor, becomes a dynamical variable related to the spin density of matter \cite{ScKi,Hehl,review,Lor,Niko}.
Since Dirac fields couple to the connection, the spin of fermions acts like a source of torsion.
The ECSK theory also passes all tests of general relativity, because even at nuclear densities the contribution from torsion to the Einstein equations is negligibly small and both theories give indistinguishable predictions at these densities.

At extremely high densities existing in black holes and in the very early universe, the minimal spinor--torsion coupling manifests itself as gravitational repulsion, which avoids the formation of singularities from fermionic matter \cite{avert,HHK,Dir,Gas,Bre,torsion1,torsion2,torsion3,Mag}.
Accordingly, the singular big bang is replaced by a nonsingular big bounce, before which the universe was contracting \cite{Kuc,Gar,torsion1}.
In addition to eliminating the initial singularity, this scenario solves the flatness and horizon problems in cosmology \cite{torsion1,torsion2,torsion3,Mag}.
Cosmic inflation, which is supported by the measurements of the inhomogeneities of the cosmic microwave background radiation \cite{Ric}, also solves the flatness and horizon problems, but it requires additional matter fields with specific conditions on their form and does not address the big bang singularity \cite{infl}.
Torsion therefore provides the simplest and most natural mechanism that solves these three major problems of the standard big bang cosmology.
The ECSK theory may also solve the problem of divergent integrals in quantum field theory by providing fermions with spatial extension and thus introducing an effective ultraviolet cutoff for their propagators \cite{Dir}.

The contraction of the universe before the big bounce must be caused by a physical initial condition.
The general features of nonsingular universes and bouncing cosmologies have been reviewed in the literature \cite{prop}.
Such a contraction could correspond to gravitational collapse of matter inside a newly formed black hole existing in another universe \cite{torsion1,Pat}.
To describe gravitational collapse of a dustlike medium (Tolman--Bondi collapse) in general relativity, one can choose a reference system that is both synchronous and comoving \cite{LL2}.
Accordingly, each spatial point in the interior of a black hole locally evolves toward the singularity as an independent, spatially homogeneous, and isotropic universe \cite{Lor,LL2}.
Numerical analysis of generic gravitational collapse in general relativity shows that spatial derivatives of the metric and curvature tensors become negligible in the dynamics, and each spatial point in the interior of a black hole locally evolves toward the singularity as an independent, spatially homogeneous universe \cite{num}.
This evolution looks like the Belinskii--Khalatnikov--Lifshitz oscillatory dynamics \cite{BKL}.

In general relativity, the singularity forms before the collapse has completed \cite{LL2,HoVi}.
The infalling matter and fields make the interior of the black hole a very dynamical spacetime where the inhomogeneities and anisotropies are very large.
During the collapse, multiple trapped null surfaces with a very complicated structure can form in the interior and fold in on themselves \cite{HoVi}.
In the ECSK theory, we expect that each spatial point evolves toward a state of an extremely high but finite density and curvature.
At such a state, the local contraction ends, the matter undergoes a bounce, and the local expansion begins.
Accordingly, multiple (at least two) dynamical wormhole throats form \cite{HoVi}.
Quantum effects in the presence of an extremely strong gravitational field cause an intense particle production, which creates an enormous amount of mass \cite{Zel} without changing the total energy (matter plus gravitational field) in the black hole \cite{torsion4}.
This creation may increase the existing inhomogeneities and anisotropies in the interior \cite{tech}.

The dynamics of spacetime in a black hole is very complex, but we conjecture that eventually the wormholes will merge into one wormhole and the outermost trapped surface can be regarded as the event horizon.
Asymptotically, the throat of this wormhole and the event horizon would coincide.
The entire interior of a black hole would then become a new, closed universe whose expansion is not visible from the outside of the black hole because of an infinite redshift at its event horizon.
Such a universe would be initially inhomogeneous and anisotropic.
Its expansion would eventually stop because of the closed geometry \cite{LL2}.
The universe in a black hole would then contract until it reaches a minimum radius and maximum density, undergoes a bounce, and expands again.
Subsequent contracting phases and bounces could make it more homogeneous and isotropic.
The spin tensor could also contribute to the evolution of the shear and twist tensors in the Raychaudhuri equation on a timescale to allow a homogeneous and isotropic universe to form \cite{tech}.
This universe can be thought of as a three-dimensional analog of the two-dimensional surface of a sphere \cite{torsion1}.

When the universe in a black hole creates sufficient amounts of mass, it reaches the size at which the cosmological constant becomes dominant and the expansion proceeds to infinity \cite{Lor}.
The boundary of the black hole (event horizon) becomes an Einstein--Rosen bridge (wormhole) connecting this universe with the outer universe in which the black hole exists.
Every astrophysical black hole candidate may thus be a wormhole to a new universe on the other side of its event horizon.
Accordingly, our universe may be the interior of a black hole existing in another universe and a part of a multiverse.
This scenario could also explain the arrow of time and the black hole information paradox \cite{torsion1,torsion4}.

This paper is structured as follows. Section 2 briefly describes the ECSK theory of gravity.
Section 3 briefly describes a spin fluid.
Section 4 presents the cosmology of the early universe in the presence of spin and torsion.
The next sections constitute the novel part of the paper.
Section 5 analyzes the dynamics of a relativistic universe in a black hole without particle production.
For simplicity, we consider a homogeneous and isotropic universe.
Section 6 adds quantum particle production caused by an extremely high curvature and analyzes its effects on the dynamics of a universe in a black hole.
Section 7 shows that inflationary expansion can be derived as a special case of the dynamics described in section 6.
Cosmic inflation can thus originate from the ECSK theory of gravity with particle production, without requiring additional matter fields. The results are briefly discussed in section 8.\\

{\bf 2. ECSK theory of gravity}\\ \\
The affine connection $\Gamma^{k}_{ij}$ in the ECSK theory of gravity is not symmetric and has the antisymmetric part, the torsion tensor $S^k_{\phantom{k}ij}=(\Gamma^{k}_{ij}-\Gamma^{k}_{ji})/2$ \cite{ScKi,Hehl,review,Lor,Niko,HHK,Dir,Schr,HO}.
We use the notation of \cite{Niko}.
The curvature tensor is given by $R^i_{\phantom{i}mjk}=\partial_{j}\Gamma^{i}_{mk}-\partial_{k}\Gamma^{i}_{mj}+\Gamma^{i}_{lj}\Gamma^{l}_{mk}-\Gamma^{i}_{lk}\Gamma^{l}_{mj}$, and its contraction gives the Ricci tensor $R_{ik}=R^j_{\phantom{j}ijk}$.
The metricity condition $g_{ij;k}=0$, where a semicolon denotes the covariant derivative with respect to $\Gamma^{k}_{ij}$, gives $\Gamma^{k}_{ij}=\{^{k}_{ij}\}+C^k_{\phantom{k}ij}$, where $\{^{k}_{ij}\}=(1/2)g^{km}(g_{mi,j}+g_{mj,i}-g_{ij,m})$ are the Christoffel symbols of the metric tensor $g_{ik}$ and $C^i_{\phantom{i}jk}=S^i_{\phantom{i}jk}+2S_{(jk)}^{\phantom{(jk)}i}$ is the contortion tensor.
The curvature tensor can be decomposed as $R^i_{\phantom{i}klm}=P^i_{\phantom{i}klm}+C^i_{\phantom{i}km:l}-C^i_{\phantom{i}kl:m}+C^j_{\phantom{j}km}C^i_{\phantom{i}jl}-C^j_{\phantom{j}kl}C^i_{\phantom{i}jm}$, where $P^i_{\phantom{i}klm}$ is the Riemann tensor (the curvature tensor constructed from the Levi-Civita connection $\{^{k}_{ij}\}$) and a colon denotes the covariant derivative with respect to $\{^{k}_{ij}\}$.

The ECSK theory of gravity is based on the Lagrangian density of the gravitational field, which is proportional to the Ricci curvature scalar $R=R_{ik}g^{ik}$, similarly to the metric general relativity.
The field equations are obtained from the total action for the gravitational field and matter, $I=(1/c)\int(-R\sqrt{-g}/(2\kappa)+\mathfrak{L}_\textrm{m})d^4x$, where $\mathfrak{L}_\textrm{m}$ is the Lagrangian density of matter, $g=\mbox{det}(g_{ik})$, and $\kappa=8\pi G/c^4$, with respect to the metric and torsion tensors.
Varying the action with respect to the torsion gives the Cartan field equations that relate algebraically the torsion of spacetime to the canonical spin tensor of matter $s^{ijk}=2(\delta\mathfrak{L}_\textrm{m}/\delta C_{ijk})/\sqrt{-g}$ \cite{ScKi,Hehl,review,Lor,Niko,HHK,Dir}:
\begin{equation}
S_{jik}-S_i g_{jk}+S_k g_{ji}=-\frac{1}{2}\kappa s_{ikj},
\label{Car}
\end{equation}
where $S_i=S^k_{\phantom{k}ik}$.
Varying the action with respect to the metric and using equation (\ref{Car}) gives the Einstein field equations that relate the curvature of spacetime to the canonical energy--momentum tensor of matter $\sigma_{ij}$:
\begin{equation}
R_{ik}-\frac{1}{2}Rg_{ik}=\kappa\sigma_{ki}.
\label{Ein}
\end{equation}
The symmetric, dynamical energy--momentum tensor $T_{ik}=2(\delta\mathfrak{L}_\textrm{m}/\delta g^{ik})/\sqrt{-g}$ \cite{LL2,Schr} is related to the canonical energy--momentum tensor by $T_{ik}=\sigma_{ik}-(1/2)(\nabla_j-2S^l_{\phantom{l}jl})(s_{ik}^{\phantom{ik}j}-s_{k\phantom{j}i}^{\phantom{k}j}+s^j_{\phantom{j}ik})$, where $\nabla_k$ denotes the covariant derivative with respect to $\Gamma^{k}_{ij}$.

The Einstein and Cartan equations can be combined to give \cite{review,Niko,HHK,Dir}
\begin{equation}
G^{ik}=\kappa T^{ik}+\frac{1}{2}\kappa^2\biggl(s^{ij}_{\phantom{ij}j}s^{kl}_{\phantom{kl}l}-s^{ij}_{\phantom{ij}l}s^{kl}_{\phantom{kl}j}-s^{ijl}s^k_{\phantom{k}jl}+\frac{1}{2}s^{jli}s_{jl}^{\phantom{jl}k}+\frac{1}{4}g^{ik}(2s^{\phantom{j}l}_{j\phantom{l}m}s^{jm}_{\phantom{jm}l}-2s^{\phantom{j}l}_{j\phantom{l}l}s^{jm}_{\phantom{jm}m}+s^{jlm}s_{jlm})\biggr),
\label{GR}
\end{equation}
where $G_{ik}=P_{ik}-(1/2)Pg_{ik}$ is the Einstein tensor of general relativity constructed from the contractions of the Riemann tensor, $P_{ik}=P^j_{\phantom{j}ijk}$ and $P=P_{ik}g^{ik}$.
The second term on the right-hand side of equation (\ref{GR}) is the correction to the curvature of spacetime from the spin.
The spin tensor also appears in $T_{ik}$ because $\mathfrak{L}_\textrm{m}$ depends on torsion.
The contributions from the spin tensor to the right-hand side of the Einstein equations are significant only at extremely high densities, on the order of the Cartan density \cite{Dir}.
Below this density, the predictions of the ECSK theory do not differ from the predictions of the metric general relativity and reduce to them in vacuum, where torsion vanishes.\\

{\bf 3. Spin fluid}\\ \\
Quarks and leptons, which compose all stars, are fermions described in relativistic quantum mechanics by the Dirac equation.
Since Dirac fields couple minimally to the torsion tensor, the torsion of spacetime at microscopic scales does not vanish in the presence of fermions \cite{ScKi,Niko,Dir}.
A correct coupling of fermions to torsion should treat spinors as anticommuting fields \cite{Cha}.
At macroscopic scales, such particles can be averaged and described as a spin fluid \cite{Niko,avert,HHK,Gas,Bre,torsion1,Kuc}.
Even if the spin orientation of particles is random, the terms quadratic in the spin tensor in equation (\ref{GR}) do not vanish.
These terms are significant only at densities of matter much higher than the density of nuclear matter because of the factor $\kappa^2$.

The Bianchi identities, $R^i_{\phantom{i}n[jk;l]}=2R^i_{\phantom{i}nm[j}S^m_{\phantom{m}kl]}$ and $R^m_{\phantom{m}[jkl]}=-2S^m_{\phantom{m}[jk;l]}+4S^m_{\phantom{m}n[j}S^n_{\phantom{n}kl]}$, together with the Einstein and Cartan field equations, give the conservation laws for the canonical energy--momentum and spin tensors: $T^{ij}_{\phantom{ij}:j}=C_{jk}^{\phantom{jk}i}T^{jk}+(1/2)s_{klj}R^{klji}$ and $s_{ij\phantom{k};k}^{\phantom{ij}k}-2S_k s_{ij}^{\phantom{ij}k}=T_{ij}-T_{ji}$ \cite{ScKi,Hehl,review,Lor,Niko,HHK}.
Using the Papapetrou method of multipole expansion for these laws \cite{Pap} leads to the formulas for the macroscopic canonical energy--momentum and spin tensors of a spin fluid in the point-particle approximation \cite{Niko,Dir}.
The canonical energy--momentum tensor of a spin fluid is given by \cite{HHK}
\begin{equation}
\sigma_{ij}=c\Pi_i u_j-p(g_{ij}-u_i u_j),
\end{equation}
and its spin tensor by
\begin{equation}
s_{ij}^{\phantom{ij}k}=s_{ij}u^k,\,\,\,s_{ij}u^j=0,
\end{equation}
where $\Pi_i$ is the four-momentum density of the fluid, $u^i$ its four-velocity, $s_{ij}$ its spin density, and $p$ its pressure.
Substituting the macroscopic tensors into equation (\ref{GR}) gives \cite{HHK,Gas,Bre,torsion1,Kuc}
\begin{equation}
G^{ij}=\kappa\Bigl(\epsilon-\frac{1}{4}\kappa s^2\Bigr)u^i u^j-\kappa\Bigl(p-\frac{1}{4}\kappa s^2\Bigr)(g^{ij}-u^i u^j)-\frac{1}{2}\kappa(\delta^l_k+u_k u^l)(s^{ki}u^j+s^{kj}u^i)_{:l},
\label{com}
\end{equation}
where $\epsilon=c\Pi_i u^i$ is the rest energy density of the fluid, and
\begin{equation}
s^2=\frac{1}{2}s_{ij}s^{ij}>0
\end{equation}
is the square of the spin density.

If the spin orientation of particles in a spin fluid is random, then the last term on the right-hand side of equation (\ref{com}) vanishes after averaging.
Thus, the Einstein--Cartan equations for such a spin fluid are equivalent to the Einstein equations for a perfect fluid with the effective energy density $\epsilon-\kappa s^2/4$ and the effective pressure $p-\kappa s^2/4$ \cite{HHK,Gas,Bre,torsion1,Kuc}.
The square of the spin density for a fluid consisting of fermions with no spin polarization is given by \cite{Gas,Bre,NuPo}
\begin{equation}
s^2=\frac{1}{8}(\hbar cn_\textrm{f})^2,
\end{equation}
where $n_\textrm{f}$ is the number density of fermions.
The effective energy density and pressure of a spin fluid are thus given by
\begin{equation}
\tilde{\epsilon}=\epsilon-\alpha n_\textrm{f}^2,\quad\tilde{p}=p-\alpha n_\textrm{f}^2,
\label{eff}
\end{equation}
where $\alpha=\kappa(\hbar c)^2/32$.\\

{\bf 4. Friedmann equations with torsion}\\ \\
A closed, homogeneous, and isotropic universe is described by the Friedmann--Lema\^{i}tre--Robertson--Walker (FLRW) metric, which, in the isotropic spherical coordinates, is given by $ds^2=c^2 dt^2-a^2(t)(1+kr^2/4)^{-2}(dr^2+r^2 d\theta^2+r^2\sin^2\theta d\phi^2)$, where $a(t)$ is the scale factor and $k=1$ \cite{Lor,LL2}.
In the comoving frame of reference, in which the four-velocity $u^i$ of the cosmological spin fluid satisfies $u^0=1$ and $u^\alpha=0$ ($\alpha$ denotes spatial indices), the Einstein field equations (\ref{com}) for this metric become the Friedmann equations (the cosmological constant is negligible in the early universe) \cite{Gas,Bre,torsion1,Kuc}:
\begin{eqnarray}
& & \frac{{\dot{a}}^2}{c^2}+k=\frac{1}{3}\kappa(\epsilon-\alpha n_\textrm{f}^2)a^2, \label{Fri1} \\
& & \frac{{\dot{a}}^2+2a\ddot{a}}{c^2}+k=-\kappa(p-\alpha n_\textrm{f}^2)a^2,
\label{Fri2}
\end{eqnarray}
where a dot denotes the differentiation with respect to the cosmic time $t$.
Differentiating equation (\ref{Fri1}) multiplied by $a$ with respect to $t$ and combining it with equation (\ref{Fri2}) multiplied by $\dot{a}$ give a conservation law
\begin{equation}
\frac{d}{dt}\bigl((\epsilon-\alpha n_\textrm{f}^2)a^3\bigr)+(p-\alpha n_\textrm{f}^2)\frac{d}{dt}(a^3)=0.
\label{law}
\end{equation}

The spin fluid in the early universe is formed by ultrarelativistic matter in kinetic equilibrium, for which $\epsilon(T)=h_\star T^4$, $p(T)=\epsilon(T)/3$, and $n(T)=h_n T^3$, where $T$ is the temperature of the universe and $\zeta$ is the Riemann zeta function, and we defined $h_\star=(\pi^2/30)g_\star k_\textrm{B}^4/(\hbar c)^3$ and $h_n=(\zeta(3)/\pi^2)g_n k_\textrm{B}^3/(\hbar c)^3$ \cite{Ric}.
The effective numbers of thermal degrees of freedom are $g_\star(T)=g_\textrm{b}(T)+(7/8)g_\textrm{f}(T)$ and $g_n(T)=g_\textrm{b}(T)+(3/4)g_\textrm{f}(T)$, where $g_\textrm{b}=\sum_i g_i$ is summed over relativistic bosons, $g_\textrm{f}=\sum_i g_i$ is summed over relativistic fermions, and $g_i$ is the number of the spin states for each particle species $i$.
Accordingly, $n_\textrm{f}(T)=h_{n\textrm{f}}T^3$ and $h_{n\textrm{f}}=(\zeta(3)/\pi^2)(3/4)g_\textrm{f}(T)(k_\textrm{B}T)^3/(\hbar c)^3$.
Substituting these functions into equation (\ref{law}) gives
\begin{equation}
\Bigl(\frac{\dot{a}}{a}+\frac{\dot{T}}{T}\Bigr)\Bigl(1-\frac{3\alpha h_{n\textrm{f}}^2}{2h_\star}T^2\Bigr)=0,
\label{diff}
\end{equation}
since $g_\textrm{b}$ and $g_\textrm{f}$ are constant in the range of $T$ where torsion is significant.
Relation (\ref{diff}) is satisfied for all values of $T$ if
\begin{equation}
\frac{d}{dt}(aT)=0.
\label{con}
\end{equation}
Integrating equation (\ref{con}) gives
\begin{equation}
a=\frac{a_\textrm{r} T_\textrm{r}}{T},
\label{prop}
\end{equation}
where $a_\textrm{r}$ is the scale factor at a reference temperature $T_\textrm{r}$.
Relation (\ref{prop}) between $a$ and $T$ does not depend on the presence of spin \cite{torsion1}.
The energy density of particles in kinetic equilibrium scales like $\epsilon\sim a^{-4}$, and their number density scales like $n\sim a^{-3}$.
Accordingly, the number of particles $N\propto na^3$ in the universe does not change if equation (\ref{con}) is satisfied.\\

{\bf 5. Oscillatory universe without particle production}\\ \\
When a trapped null surface forms in a black hole \cite{HoVi}, the region within this surface is causally connected and mathematically equivalent to a new, closed universe \cite{torsion1,num}.
We will also assume, for simplicity, that such a universe is already homogeneous and isotropic \cite{HoVi}.
In a black hole, fermions composing the spin fluid have energies much greater than their rest energies.
These particles are thus described by an ultrarelativistic barotropic equation of state, $p=\epsilon/3$, as for radiation.
If we assume that the universe begins to contract when $a=a_i$ and $\dot{a}=0$, where $\alpha n_\textrm{f}^2\ll\epsilon$, then equation (\ref{Fri1}) gives $\kappa\epsilon a_i^2/3=1$.
Accordingly, the initial temperature is given by
\begin{equation}
T_i=\Bigl(\frac{3}{\kappa h_\star a_i^2}\Bigr)^{1/4}.
\end{equation}
If no particles are produced, then we can use equation (\ref{con}).
The reference values in equation (\ref{prop}) can be $a_\textrm{r}=a_i$ and $T_\textrm{r}=T_i$.
The dynamics of the new universe is described by equations (\ref{prop}) and (\ref{Fri1}), which can be written as
\begin{equation}
\frac{{\dot{a}}^2}{c^2}+1=\frac{1}{3}\kappa(h_\star T^4-\alpha h_{n\textrm{f}}^2 T^6)a^2
\label{osc}
\end{equation}
or
\begin{equation}
\frac{{\dot{a}}^2}{c^2}+1=\frac{1}{3}\kappa\Bigl(h_\star\frac{T_i^4 a_i^4}{a^2}-\alpha h_{n\textrm{f}}^2\frac{T_i ^6 a_i^6}{a^4}\Bigr).
\end{equation}
The Friedmann equations (\ref{Fri1}) and (\ref{Fri2}) also give
\begin{equation}
\frac{\ddot{a}}{c^2 a}=-\frac{1}{3}\kappa\epsilon+\frac{2}{3}\kappa\alpha n_\textrm{f}^2=-\frac{\kappa}{3}(h_\star T^4-2\alpha h_{n\textrm{f}}^2 T^6).
\label{acc}
\end{equation}

As the scale factor $a$ decreases, the temperature $T$ increases.
The spin--torsion contribution to the energy density is negative and generates gravitational repulsion, which becomes stronger as $T$ increases.
Since this contribution scales with $T$ faster ($T\sim T^6$) than $\epsilon\sim T^4$, $\dot{a}$ eventually reaches zero and the universe undergoes a nonsingular bounce.
At a bounce, where we can neglect $k$, equation (\ref{Fri1}) gives $\epsilon-\alpha n_\textrm{f}^2=0$.
Accordingly, the maximum temperature is given by
\begin{equation}
T_\textrm{max}=\Bigl(\frac{h_\star}{\alpha h_{n\textrm{f}}^2}\Bigr)^{1/2}=\sqrt{\frac{2\pi^5}{15}}\frac{g_\star^{1/2}}{\zeta(3)(3/4)g_\textrm{f}}T_\textrm{P},
\label{Tmax}
\end{equation}
where $T_\textrm{P}$ is the Planck temperature.
The temperature given by equation (\ref{Tmax}) depends only on the numbers of thermal degrees of freedom.
Relation (\ref{prop}) determines the minimum scale factor at a bounce of the universe in a black hole:
\begin{equation}
a_\textrm{min}=\frac{a_i T_i}{T_\textrm{max}}.
\end{equation}
The universe is accelerating if $\ddot{a}>0$, which is equivalent, using equation (\ref{acc}), to $T>T_\textrm{max}/\sqrt{2}$ ($a<\sqrt{2}a_\textrm{min}$).
After a bounce, the universe accelerates until $T=T_\textrm{max}/\sqrt{2}$ ($a=\sqrt{2}a_\textrm{min}$).
The universe continues to expands with $\ddot{a}<0$ until it reaches its maximum scale factor (a crunch) at $a=a_i$, after which another contraction begins.
This expansion looks like the time reversal of the contraction from $a=a_i$ to $a=a_\textrm{min}$.
The universe is therefore oscillatory: its scale factor oscillates between the minimum and maximum scale factors \cite{Kuc}.
The period of the oscillation from a bounce to a crunch and back is equal to
\begin{equation}
\Delta t=2\int_{a_\textrm{min}}^{a_i}\frac{da}{\dot{a}}=\frac{2}{c}\int_{a_\textrm{min}}^{a_i}\frac{da}{(\kappa\tilde{\epsilon}a^2/3-1)^{1/2}},
\end{equation}
where $\tilde{\epsilon}=h_\star(a_i T_i/a)^4-\alpha h_{n\textrm{f}}^2(a_i T_i/a)^6$.

The scale factor depends on the time according to
\begin{equation}
\Bigl(a^2-\frac{\alpha h_{n\textrm{f}}^2 a_i^2 T_i^2}{h_\star}-\frac{3ka^4}{\kappa h_\star a_i^4 T_i^4}\Bigr)^{-1/2}a^2 da=\Bigl(\frac{\kappa h_\star}{3}\Bigr)^{1/2}(a_i T_i)^2 cdt.
\label{dyn}
\end{equation}
For $a\ll a_i$, the term with $k$ in equation (\ref{dyn}) can be neglected.
If we define $x=a/a_\textrm{min}$ and choose $t=0$ at $a=a_\textrm{min}$, then integrating equation (\ref{dyn}) gives \cite{torsion1}
\begin{equation}
\frac{t}{\tau}=\int_1^x\frac{x^2 dx}{(x^2-1)^{1/2}}=\frac{x}{2}\sqrt{x^2-1}+\frac{1}{2}\mbox{ln}|x+\sqrt{x^2-1}|,
\end{equation}
where
\begin{equation}
\tau=\frac{\alpha h_{n\textrm{f}}^2}{c}\Bigl(\frac{3}{\kappa h_\star^3}\Bigr)^{1/2}=\frac{45}{4}\sqrt{\frac{5}{\pi^{13}}}\frac{\zeta^2(3)((3/4)g_\textrm{f})^2}{g_\star^{3/2}}t_\textrm{P}
\end{equation}
is the characteristic time of the torsion-dominated phase of the expansion.
This time is on the order of the Planck time $t_\textrm{P}$ and depends only on the numbers of thermal degrees of freedom.

The Hubble parameter $H=\dot{a}/a$ is determined by
\begin{equation}
H=\pm c\Bigl(\frac{1}{3}\kappa(h_\star T^4-\alpha h_{n\textrm{f}}^2 T^6)-\frac{k}{a^2}\Bigr)^{1/2},
\end{equation}
where a plus sign corresponds to expansion and a minus sign corresponds to contraction.
The magnitude of $H$ has a maximum $|H|_\textrm{max}=\sqrt{4/27}\tau^{-1}$ at $T=\sqrt{2/3}T_\textrm{max}$ (at $a=\sqrt{3/2}a_\textrm{min})$.
The total density parameter $\Omega=\tilde{\epsilon}/(\rho_c c^2)$, where $\rho_c=3H^2/(\kappa c^4)$ is the critical density, is determined by equation (\ref{Fri1}): $\Omega=1+kc^2/\dot{a}^2$, which leads to
\begin{equation}
\Omega=1+\frac{3}{\kappa(h_\star T^4-\alpha h_{n\textrm{f}}^2 T^6)a^2}=1+\frac{3}{\kappa(h_\star T^2-\alpha h_{n\textrm{f}}^2 T^4)a_i^2 T_i^2}.
\label{dens}
\end{equation}
The density parameter has a minimum $\Omega_\textrm{min}=1+4c\tau/a_i$ at $T=\sqrt{1/2}T_\textrm{max}$ (at $a=\sqrt{2}a_\textrm{min})$.
A closed universe contains $N\approx(\dot{a}/c)^3=(\Omega-1)^{-3}$ causally disconnected regions.
Their maximum number is thus $N_\textrm{max}\approx(a_i/l_\textrm{P})^3$, where $l_\textrm{P}$ is the Planck length.
As the universe expands from $a_\textrm{min}$ to $\sqrt{2}a_\textrm{min}$, $\Omega$ rapidly decreases from infinity to $\Omega_\textrm{min}$, which appears to be tuned to 1 to a high precision since $c\tau\ll a_i$ \cite{torsion1}.
As the universe expands from $\sqrt{2}a_\textrm{min}$ to $a_i$, $\Omega$ increases to infinity.

If we assume that only the known standard-model particles exist, then $g_\textrm{b}=28$ and $g_\textrm{f}=90$ \cite{Ric}.
For these values, we find
\begin{eqnarray}
& & T_\textrm{max}=1.15\times 10^{32}\,\mbox{K},\quad\tau=4.75\times 10^{-45}\,\mbox{s}, \nonumber \\
& & |H|_\textrm{max}=8.1\times 10^{43}\,\mbox{s}^{-1},\quad\Omega_\textrm{min}-1=5.7\times 10^{-36},\quad N_\textrm{max}\approx 10^{52}.
\label{values}
\end{eqnarray}
An extremely small magnitude of $\Omega_\textrm{min}-1$ solves the flatness problem.
An extremely large $N$ solves the horizon problem.
A typical stellar black hole has the Schwarzschild radius $a_i=10^4$ m.
For this value, we find
\begin{equation}
T_i=1.38\times 10^{12}\,\mbox{K},\quad a_\textrm{min}=1.19\times 10^{-16}\,\mbox{m}.
\end{equation}

The volume of a closed universe is given by $V=2\pi^2 a^3$ \cite{LL2}.
The conservation law given by equation (\ref{law}) can thus be written in the form of the first law of thermodynamics \cite{Gar}:
\begin{equation}
dE+pdV=TdS=12\pi^2\alpha h_{n\textrm{f}}^2 a^2 T^5 d(aT),
\label{ent}
\end{equation}
where $E=\epsilon V$ is the thermal energy of the universe and $S$ its entropy.
This law also leads to a lower limit for the scale factor of the universe \cite{Gar}.
Relations (\ref{con}) and (\ref{ent}) show that the entropy of a cyclic universe in a black hole without particle production is constant.\\

{\bf 6. Universe with particle production}\\ \\
A closed universe in a black hole without particle production is oscillatory.
It does not reach the size and mass of the observed universe.
To identify our universe with a universe in a black hole, we must include quantum effects in curved spacetime, which are responsible for particle production by the gravitational field \cite{Zel}.

An initial, accelerated expansion of a closed universe after the big bounce defines the torsion-dominated era.
As the universe expands, the spin--torsion term $\alpha n_\textrm{f}^2$ in equation (\ref{Fri1}) decreases faster than $\epsilon$.
The universe begins to decelerate and enters the radiation-dominated era.
Eventually, the matter in the universe becomes nonrelativistic.
At the matter--radiation equality, the energy density of nonrelativistic matter exceeds the energy density of radiation and the universe enters the matter-dominated era.
The universe cannot expand to infinity unless it enters another phase of acceleration.
The simplest source of such an acceleration is a cosmological constant $\Lambda$, which modifies the Lagrangian density of the gravitational field: $\mathfrak{L}_\textrm{g}=-(R+2\Lambda)\sqrt{-g}/(2\kappa)$.
The cosmological constant does not change the Cartan equations but adds a term to the Einstein equations (\ref{Ein}): $R_{ik}-(1/2)Rg_{ik}=\kappa\sigma_{ki}+\Lambda g_{ik}$.
The Friedmann equation (\ref{Fri1}) in the radiation-dominated era becomes
\begin{equation}
\frac{{\dot{a}}^2}{c^2}+1=\frac{1}{3}\kappa\epsilon a^2+\frac{1}{3}\Lambda a^2.
\label{Fri3}
\end{equation}
In the matter-dominated era, the energy density is dominated by the mass density of nonrelativistic matter $\rho$: $\epsilon\approx\rho c^2$.
The mass of the nonrelativistic matter in the universe is equal to $M_\textrm{univ}=\rho V$.
Accordingly, the Friedmann equation (\ref{Fri1}) in the matter-dominated era is
\begin{equation}
\frac{{\dot{a}}^2}{c^2}+1=\frac{D}{a}+\frac{1}{3}\Lambda a^2,
\label{Fri4}
\end{equation}
where $D=4GM_\textrm{univ}/(3\pi c^2)$.

A closed universe in a black hole expands to infinity if \cite{Lor}
\begin{equation}
D>\frac{2}{3\sqrt{\Lambda}}.
\label{inf}
\end{equation}
At the matter--radiation equality, where $a=a_\textrm{eq}$ and $T=T_\textrm{eq}$, we have $(1/3)\kappa\epsilon_\textrm{eq} a_\textrm{eq}^2=D/a_\textrm{eq}$ or
\begin{equation}
D=\frac{1}{3}\kappa h_\star T_\textrm{eq}^4 a_\textrm{eq}^3.
\end{equation}
Using equation (\ref{prop}), this constant can be written as
\begin{equation}
D=\frac{1}{3}\kappa h_\star T_\textrm{eq}\tilde{a}_i^3 T_i^3=\frac{1}{3}\kappa h_\star T_\textrm{eq}\Bigl(\frac{\tilde{a}_i}{a_i}\Bigr)^3(a_i T_i)^3,
\end{equation}
where $\tilde{a}_i$ is the scale factor at $T=T_i$ in the expanding phase.
Without particle production, $\tilde{a}_i=a_i$.
With particle production, $\tilde{a}_i>a_i$.
Since $\tilde{a}_i$ and $a_i$ correspond to the same temperature, the ratio $\tilde{a}_i/a_i$ represents the expansion of the universe due to particle production.
Relation (\ref{inf}) shows that
\begin{equation}
\frac{\tilde{a}_i}{a_i}>\Bigl(\frac{2}{\kappa h_\star T_\textrm{eq}(a_i T_i)^3\sqrt{\Lambda}}\Bigr)^{1/3}.
\end{equation}
The value of $T_\textrm{eq}$ depends on the number ratio of photons to massive particles and on the masses of the particles.
In our universe, $T_\textrm{eq}=8820$ K and $\Lambda/\kappa=5.24\times 10^{-10}$ Pa \cite{Ric}, which gives
\begin{equation}
\frac{\tilde{a}_i}{a_i}>10^{10}.
\label{cond}
\end{equation}
Accordingly, equations (\ref{dens}) and (\ref{values}) lead to
\begin{equation}
\tilde{\Omega}_\textrm{min}-1=(\Omega_\textrm{min}-1)\Bigl(\frac{a_i}{\tilde{a}_i}\Bigr)^2<10^{-55}.
\end{equation}
This value solves both the flatness and horizon problems \cite{torsion1}.

If equation (\ref{inf}) is not satisfied, then the universe in a black hole eventually reaches a crunch at which it stops expanding and starts contracting.
The contraction lasts until the universe reaches the next bounce at $T=T_\textrm{max}$ and $a>a_\textrm{min}$.
After this bounce, the next expanding phase begins.
The universe is more isotropic and has a larger scale factor at $T=T_\textrm{eq}$ than in the previous expanding phase.
Accordingly, it has a larger value of $D$.
If this new value satisfies equation (\ref{inf}), then the universe expands to infinity.
If not, the universe has another cycle of contraction and expansion.
After a finite number of cycles, $D$ satisfies equation (\ref{inf}) and the universe expands indefinitely.
The last bounce can be regarded as the big bang, or rather the big bounce, of this universe.

In an isotropic universe, described by the FLRW metric, an alternating gravitational field causes production of particle--antiparticle pairs \cite{prod}.
For a strong field in general relativity, the local rate of production of massive particles with spin $j=1$ is equal to
\begin{equation}
\frac{1}{\sqrt{-g}}\frac{d}{dt}(\sqrt{-g}n_1)=\frac{c}{288\pi}P^2,
\label{rate}
\end{equation}
where $n_1$ is their number density.
The local rates of production of massless particles with spin $j=1$ and particles with spins $j=0$ and $j=1/2$ are proportional to $P$ and can be neglected.
In the ECSK theory, the torsion tensor may modify equation (\ref{rate}).
We can thus write
\begin{equation}
\frac{1}{\sqrt{-g}}\frac{d}{dt}(\sqrt{-g}n_1)=cK,
\end{equation}
where $K>0$ is a scalar of dimension m$^{-4}$, constructed from the curvature, torsion, and metric tensors.
For the FLRW metric, we have
\begin{equation}
\frac{1}{a^3}\frac{d}{dt}(a^3 n_1)=cK.
\end{equation}
Using $n_1(T)=h_{n1}T^3$, where $h_{n1}=(\zeta(3)/\pi^2)g_{n1}k_\textrm{B}^3/(\hbar c)^3$ and $g_{n1}=g_{\textrm{b}1}=\sum_i g_i$ is summed over relativistic, massive bosons with spin $j=1$, we find
\begin{equation}
\frac{\dot{a}}{a}+\frac{\dot{T}}{T}=\frac{cK}{3h_{n1}T^3}.
\label{equ}
\end{equation}
If we assume that equation (\ref{Fri1}) is satisfied, then equation (\ref{Fri2}) must be modified. 
Relation (\ref{equ}) generalizes equation (\ref{con}) and describes, together with equation (\ref{osc}), the dynamics of the universe.

Particle production does not change the total energy and momentum of the matter and gravitational field in the universe \cite{torsion4}.
However, it increases the entropy in the universe.
Combining equations (\ref{ent}) and (\ref{equ}) gives
\begin{equation}
\frac{dS}{dt}=\frac{2\alpha ch_{n\textrm{f}}^2}{h_{n1}}a^3 T^2 K.
\end{equation}
The motion of matter through the event horizon is unidirectional, and thus it defines the arrow of time in the universe in a black hole.
This arrow is also entropic: although black holes are states of maximum entropy in the frame of reference of outside observers, new universes expanding inside black holes continue to increase entropy.

For the FLRW metric, the Weyl tensor vanishes and the two independent invariants of dimension m$^{-4}$ constructed from the Riemann and metric tensors are $P^2$ and $P_{ik}P^{ik}$.
Using equation (\ref{com}) without the last term on the right-hand side and equation (\ref{eff}) gives
\begin{eqnarray}
& & P=-\kappa(\tilde{\epsilon}-3\tilde{p})=-2\kappa\alpha n_\textrm{f}^2, \nonumber \\
& & P_{ik}P^{ik}=\kappa^2(\tilde{\epsilon}^2+3\tilde{p}^2).
\end{eqnarray}
At a bounce, where $\dot{a}=\tilde{\epsilon}=0$ and $T=T_\textrm{max}$, the rate of production of particles (and thus $K$) should vanish to avoid $\dot{T}>0$ in equation (\ref{equ}).
Otherwise, the increase of $T$ would lead to $T>T_\textrm{max}$ and $\dot{a}^2<0$ in equation (\ref{osc}).
A scalar, vanishing for $\tilde{\epsilon}=0$ and constructed from the Riemann and metric tensors, is proportional to $K=P^2-3P_{ik}P^{ik}$.
Including the torsion tensor can give other forms for $K$.
Ultimately, $K$ should be derived from quantum field theory in the Riemann--Cartan spacetime of the ECSK theory of gravity.

The simplest form of $K$ that vanishes at a bounce and has the same dimension as $P^2$ is
\begin{equation}
K=\beta(\kappa\tilde{\epsilon})^2,
\label{prod}
\end{equation}
where $\beta>0$ is a nondimensional constant.
Near a bounce, where we can neglect $k$, equation (\ref{Fri1}) can be written as
\begin{equation}
\Bigl(\frac{\dot{a}}{ca}\Bigr)^2=\frac{1}{3}\kappa\tilde{\epsilon}.
\label{Fri5}
\end{equation}
Combining equations (\ref{equ}), (\ref{prod}), and (\ref{Fri5}) gives
\begin{equation}
\frac{\dot{a}}{a}\Bigl[1-\frac{3\beta}{c^3 h_{n1}T^3}\Bigl(\frac{\dot{a}}{a}\Bigr)^3\Bigr]=-\frac{\dot{T}}{T}.
\label{ineq}
\end{equation}
The signs of $\dot{a}$ and $\dot{T}$ must be opposite to avoid an indefinitely long, exponential increase of the scale factor.
This condition is satisfied during a contracting phase ($\dot{a}<0$).
During an expanding phase ($\dot{a}>0$), we must have
\begin{equation}
\frac{3\beta}{c^3 h_{n1}}\Bigl(\frac{\dot{a}}{aT}\Bigr)^3=\frac{3\beta}{h_{n1}}\Bigl(\frac{\kappa}{3}(h_\star T^2-\alpha h_{n\textrm{f}}^2 T^4)\Bigr)^{3/2}<1.
\end{equation}
The above function has a maximum at $T=T_\textrm{max}/\sqrt{2}$.
We thus find
\begin{equation}
\beta<\beta_\textrm{cr}=\frac{\sqrt{6}}{32}\frac{h_{n1}h_{n\textrm{f}}^3(\hbar c)^3}{h_\star^3}=\frac{15^3\sqrt{6}}{4\pi^{14}}\frac{\zeta^4(3)g_{\textrm{b}1}((3/4)g_\textrm{f})^3}{g_\star^3}.
\end{equation}
If we assume that only the known standard-model particles exist, then $g_{\textrm{b}1}=9$ \cite{Ric}.
For this value, we find
\begin{equation}
\beta_\textrm{cr}\approx\frac{1}{929},
\end{equation}
which is on the order of $1/(288\pi)$ in equation (\ref{rate}).\\

{\bf 7. Inflation as a special case}\\ \\
The exponential expansion of space in the early universe, known as cosmic inflation, also solves the flatness and horizon problems \cite{infl}.
During inflation, the scale factor increases according to $a\propto e^{Ht}$ with constant values of $H$ and $T$.
In addition, inflation predicts the observed spectrum of primordial density fluctuations \cite{pert}.
Standard inflation, which assumes the existence of a fundamental scalar field with a specific potential, requires fine-tuning and is geodesically incomplete in the past \cite{inc}.

If $\beta$ is slightly lesser than $\beta_\textrm{cr}$, then at $T=T_\textrm{max}/\sqrt{2}$:
\begin{equation}
\Bigl|\frac{3\beta}{c^3 h_{n1}}\Bigl(\frac{\dot{a}}{aT}\Bigr)^3\Bigr|\lesssim 1.
\label{coef}
\end{equation}
At this temperature, equation (\ref{ineq}) therefore gives
\begin{equation}
\frac{\dot{T}}{T}\approx-\frac{2\dot{a}}{a}
\end{equation}
in a contracting phase and
\begin{equation}
\dot{T}\approx 0
\end{equation}
in an expanding phase.
Accordingly, equation (\ref{equ}) in an expanding phase gives an exponential expansion:
\begin{equation}
\frac{\dot{a}}{a}\approx\frac{c\beta(\kappa\tilde{\epsilon})^2}{3h_{n1}T^3}\approx c\Bigl(\frac{1}{3}\kappa\tilde{\epsilon}\Bigr)^{1/2},
\label{exp}
\end{equation}
at a nearly constant energy density
\begin{equation}
\tilde{\epsilon}\approx\frac{h_\star^3}{8\alpha^2 h_{n\textrm{f}}^4}.
\end{equation}
As $a$ increases after a bounce, the term on the left-hand side of equation (\ref{coef}) increases from 0 to nearly 1 and $T$ decreases from $T=T_\textrm{max}$ to $T=T_\textrm{max}/\sqrt{2}$.
When $T\approx T_\textrm{max}/\sqrt{2}$, the term in equation (\ref{coef}) is nearly 1 and the universe begins to expand exponentially at a nearly constant $T$ (inflation).
An exponential expansion lasts until the term in equation (\ref{coef}) decreases significantly below 1 and $T$ continues to decrease.
The universe then starts expanding according to equation (\ref{equ}) and moves from the torsion-dominated era to the radiation-dominated era.
In a scenario with one bounce, relations (\ref{cond}) and (\ref{exp}) constrain the time interval $t_\textrm{infl}$ of the exponential expansion: $Ht_\textrm{infl}\gtrsim 23$.
This interval depends on $\tau(\beta_\textrm{cr}-\beta)^{-1}$.

For $\beta=0$, the universe in a black hole is oscillatory with an infinite number of bounces and crunches (cycles).
If $0<\beta<\beta_\textrm{cr}$, then the universe is cyclic with a finite number of cycles.
In this case, the scale factor at a crunch is greater than the scale factor at the preceding crunch.
As $\beta$ increases, the number of cycles decreases and the accelerated expansion of the universe in each cycle is closer to exponential.
If $\beta\lesssim\beta_\textrm{cr}$, then the universe has only one bounce and expands through the torsion-dominated era with a period of an inflationary phase, radiation-dominated era, matter-dominated era, and then indefinitely in the cosmological-constant era.
If $\beta\ge\beta_\textrm{cr}$, then an exponential expansion of the universe would last indefinitely (eternal inflation).
This behavior as a function of $\beta$ was confirmed numerically in \cite{Sha}.\\

{\bf 8. Remarks}\\ \\
Gravitational repulsion induced by spin and torsion, which becomes significant at extremely high densities, prevents singularities in black holes and at the big bang.
Because of this repulsion and particle production, every black hole may create a new universe on the other side of its event horizon.
The formation of trapped surfaces and wormhole throats, described in section 1, could be explored numerically in the ECSK theory.
This scenario simultaneously can explain the future of black hole interiors and the origin of the universe \cite{torsion1}.
Moreover, it naturally solves the flatness and horizon problems and can generate an inflationary phase that lasts for a finite amount of time without requiring finely tuned scalar fields, replacing $R$ in the gravitational Lagrangian density by more complicated functions, or adding other hypothetical objects.
Our universe could be a stable part of the multiverse and may contain inflating regions of spacetime (new universes) only in black holes.

Numerical analysis of this scenario \cite{Sha} showed that the dynamics of the universe is insensitive to the value of $a_i$, and thus to that of $a_\textrm{min}$, and effectively depends on the particle production coefficient $\beta$ only.
Therefore, this scenario is insensitive to the detailed circumstances leading to the formation of a closed universe in a black hole.
The same behavior is obtained if we consider the dynamics of a universe with a given value of $a_\textrm{min}$ and $T_\textrm{max}$ given by equation (\ref{Tmax}) as the initial conditions for expansion \cite{Sha}.
Equations (\ref{osc}) and (\ref{equ}) can describe the dynamics of the early universe even if we do not assume a black hole as its origin.
However, the interior of a black hole can naturally provide the conditions to create a closed universe that can reach temperatures on the order of $T_\textrm{max}$.
After such a universe is formed, spin and torsion provide the mechanism to avoid a singularity and to bounce (when the universe reaches $T_\textrm{max}$), and particle production provides the mechanism to generate inflation in the subsequent expansion.
The only problem, which is more philosophical and cannot be solved, is this: if the black hole that created our universe exists in a larger universe and that universe was created inside a black hole, then what started the entire process?

The scenario presented in this paper would be more complete if we use the spin--torsion coupling beyond the point-particle approximation of a spin fluid \cite{Dir} and combine it with the Dirac-spinor description of fermions.
We showed in \cite{torsion2} that a semiclassical spin--spin interaction, generated by the minimal coupling between the torsion tensor and Dirac spinors, averts the unphysical big bang singularity and replaces it with a cusp-like big bounce at a finite minimum scale factor, before which the universe was contracting.
In that scenario, the effective energy density and pressure were given by $\tilde{\epsilon}=\epsilon-\alpha n_\textrm{f}^2,\,\tilde{p}=p+\alpha n_\textrm{f}^2$, where $\alpha=(9/16)\kappa(\hbar c)^2$.
A similar spin--spin interaction was analyzed in \cite{KR}, obtaining the same dynamics, but the cusp-like bounce was not noted; instead, it was assumed that the compression would not be followed by the expansion.
To find whether the cusp can be smoothed out, Dirac spinors should be treated not semiclassically but as anticommuting fields \cite{Cha}.

The observed matter--antimatter asymmetry in the universe may be caused by the nonlinearity of the Dirac equation in the presence of torsion \cite{dark}.
Particle production by quantum effects in strong gravitational fields may also be affected by torsion \cite{cre}.
Primordial density fluctuations in the early universe formed in a black hole should be generated before the big bounce.
If the big bounce is followed by a period of an exponential expansion, generated by torsion and particle production, then such fluctuations should be consistent with the observed spectrum.
The consistency, for a particular range of the production rate, was recently found in \cite{Sha}.
The nature of dark matter and dark energy should be related to later epochs in the universe.
The presented scenario predicts phenomena that are similar to those in some other cosmological models such as inflationary bubbles in the multiverse \cite{infl}, cyclic universe \cite{ekp}, or holographic world \cite{hol}.
However, contrary to the other models, this scenario does not need new, hypothetical fields.
Finally, we note that the high-energy observations of photons originating from gamma-ray bursts indicate that specatime is continuous at the Planck temperature \cite{cont}, which is on the order of the temperature near a bounce.
Therefore, the classical description of the gravitational field used in this scenario may be sufficient.

I am grateful to my parents, Bo\.{z}enna Pop{\l}awska and Janusz Pop{\l}awski.
This research was supported by the University Research Scholar program at the University of New Haven.

\end{document}